# Fully Homomorphic Encryption via Affine Automorphisms

Paul Hriljac

Embry-Riddle Aeronautical University

**Abstract**

This paper describes a methodology for implementing a form of fully homomorphic encryption. The scheme applies to straight line computer programs operating on data sets composed of *n*-tuples of integers via a set of mathematical operations based on integral affine automorphisms, a classic field of study in algebraic geometry and number theory. The data is encrypted with a new type of multivariate encryption originating from the same operations. The potential resiliency of the scheme is based on the difficulty of solving or analyzing systems of nonlinear Diophantine equations. The impact of this scheme on computation and memory is analyzed and is shown to be considerably less than other current methods of fully homomorphic encryption, to an extent that could make this scheme practical for some applications. Several working examples of the scheme applied to simple programs are included.

**Introduction**

Fully homomorphic encryption (FHE) was first conjectured by Ron Rivest (see [R], [RAD]) who noticed that the RSA algorithm was homomorphic with respect to multiplication, and conjectured generalization to other types of arithmetic and logical operations. Craig Gentry (see [G1],[G2]) created the first FHE scheme. There have been a number of important additional results in this area (see [CNT],[DGHV],[GH], [GM], [M],[SS], [RS], etc.)

Generally, FHE refers to schemes for processing encrypted data without the need for ever decrypting it. A FHE scheme is composed of four procedures: key generation, encryption, decryption, and program modification that together act on families of programs and data. The program modification procedure transforms programs so that they operate on encrypted data without the need for decryption first. Typically, the encryption used is independent of the programs, for instance Gentry uses a lattice-based algorithm, then transforms programs using his method of bootstrapping.

The scheme introduced here follows a different order. The modification procedure is applied first to a program, and then the encryption and decryption procedures are created that apply to the relevant data. The modification process is performed by generating a nonlinear algebraic automorphism (a transformation consisting of a system of polynomial functions with an inverse of the same form) of the variables of the program, then rewriting the program with this transformation. Essentially a new random coordinate system is generated for the state space of the program, and all the instructions of the program are rewritten in terms of that coordinate system. Data is encrypted by expressing it in that coordinate system. Cryptoanalysis of the scheme is equivalent to discovery of the inverse of a system of nonlinear algebraic equations with integer coefficients, a problem with known difficulty (see [CLO], [LL],[Y]). The result is a scheme that transforms programs with manageable computational impact and has a resilient encryption algorithm.

**Acknowledgements**

Thanks to Jon Haass, Hisa Tsutsui, Jim Francis, Ed Poon, Matt Jaffe, David Russell, John Harrel, Tim Lewis, Jesse Walker, Drew Dean, Paul Friedrichs, Steve Huffman, Bruce Hoy and Joel Bagnal. The research in this paper was partially supported by an AFOSR STTR, USAF contract # FA8750-11-C-0141. Some computations in this paper were performed by using Maple™, a product of Maplesoft.

**Overview**

Fully Homomorphic Encryption can be thought of an encrypted data set and functions to process the data in such a way that the processing occurs without ever having to decrypt the data, until the very end, when the data has been moved to a safe resting place. An application would be a cloud-based bank. Such an organization must process their records on a server they don't control while protecting privacy and data integrity, a data-in-motion problem. We model such situations by thinking of the data set as a collection of integer-valued *n*-tuples and piecewise polynomial functions carrying these *n*-tuples to other *n*-tuples.

**Model of Computation**

Begin with a computer program $P$ expressed as a straight-line program over the integers, see [Ka]. So that we have a finite directed graph with three types of nodes (input, output, and computation), a state space $\mathbb{Z}^n$, input space $\mathbb{Z}^k$, output space $\mathbb{Z}^l$ and a collection of integer polynomial functions associated with each node of the graph. For simplicity we assume that there is only one input node and one output node. Therefore, we have:

1. $u \in \mathbb{Z}^k$, the input variables to $P$;
2. $v \in \mathbb{Z}^l$, the output variables of $P$;
3. $x \in \mathbb{Z}^n$, the state variables of $P$;
4. A polynomial input function $f_{in}: \mathbb{Z}^k \to \mathbb{Z}^n$;
5. A polynomial output function $f_{out}: \mathbb{Z}^n \to \mathbb{Z}^l$;
6. A collection of polynomial processing functions $f_{ss}^a: \mathbb{Z}^n \to \mathbb{Z}^n$
   $a \leftrightarrow computation\ nodes\ of\ the\ graph\ for\ P$

The idea behind our process is based on *integer-valued affine transformations*. These are (possibly) nonlinear, invertible polynomial mappings from integer *n*-tuples to themselves. Such mappings have been studied in algebraic geometry for a very long time. Generating these mappings are very easy but analyzing them is very hard.

**First Example**

We start with a simple but illustrative example. Consider the following Maple™ program which adds and multiplies two numbers:

*fun1* := **proc**(*U1*, *U2*)
 **local** *x*, *X1*, *X2*, *V1*, *V2*, *temp1*, *temp2*;
 *X1* := *U1*;
 *X2* := *U2*;
 *temp1* := *X1* + *X2*;
 *temp2* := *X1*·*X2*;
 *X1* := *temp1*;
 *X2* := *temp2*;
 *V1* := *X1*;
 *V2* := *X2*;
 **return**(*V1*, *V2*);
 **end proc**;

We define a pair of automorphisms of integer 2-tuples by

$$\begin{bmatrix} Y1 \\ Y2 \end{bmatrix} = \phi\left(\begin{bmatrix} X1 \\ X2 \end{bmatrix}\right) = \begin{bmatrix} -X1 - 3*X2 + 2*X2^2 \\ 2*X1 + 5*X2 - 4*X2^2 \end{bmatrix},$$
$$\begin{bmatrix} X1 \\ X2 \end{bmatrix} = \psi\left(\begin{bmatrix} Y1 \\ Y2 \end{bmatrix}\right) = \begin{bmatrix} 5*Y1 + 3*Y2 + 8*Y1^2 + 8*Y1*Y2 + 2*Y2^2 \\ -2*Y1 - Y2 \end{bmatrix}.$$

These mappings are algebraic inverses of each other. Using them one can rewrite any algebraic relation between the variable X1 and X2 in terms of the variables Y1 and Y2. We rewrite the update equations $X1 \leftarrow X1 + X2$ and $X2 \leftarrow X1*X2$ occurring above in terms of Y1 and Y2, resulting in

Y1←1664*Y1^4*Y2+1728*Y1^3*Y2^2+1536*Y1^5*Y2+1280*Y1^3*Y2^3+362*Y1^2*Y2^2
+232*Y2^4*Y1+440*Y1^3*Y2+132*Y2^3*Y1+96*Y2^5*Y1
+1920*Y1^4*Y2^2+480*Y1^2*Y2^4+896*Y1^2*Y2^3+72*Y1^2*Y2+36*Y2^2*Y1+25*Y1*Y2+512*Y1^6+640*Y1^5+48*Y1^3+22*Y1^2+200*Y1^4+7*Y2^2
+24*Y2^5+8*Y2^6+6*Y2^3+18*Y2^4-3*Y1-2*Y2;

Y2←-3328*Y1^4*Y2-3456*Y1^3*Y2^2-3072*Y1^5*Y2-2560*Y1^3*Y2^3-724*Y1^2*Y2^2-464*Y2^4*Y1-880*Y1^3*Y2-264*Y2^3*Y1-192*Y2^5*Y1
-3840*Y1^4*Y2^2-960*Y1^2*Y2^4-1792*Y1^2*Y2^3-120*Y1^2*Y2-60*Y2^2*Y1-39*Y1*Y2-1024*Y1^6-1280*Y1^5-80*Y1^3-34*Y1^2-400*Y1^4-11*Y2^2
-48*Y2^5-16*Y2^6-10*Y2^3-36*Y2^4+6*Y1+4*Y2;

Using the function $\phi$ allows for input data U1, U2 to be written in terms of Y1 and Y2. The new program is

E(fun1):= proc (U1, U2)
local X1, X2, Y1, Y2, temp1, temp2;
Y1 := -U1-3*U2+2*U2^2;

```
Y2 := 2*U1+5*U2-4*U2^2;

temp1:=1664*Y1^4*Y2+1728*Y1^3*Y2^2+1536*Y1^5*Y2+1280*Y1^3*Y2^3+362*Y1^2*Y
2^2+232*Y2^4*Y1+440*Y1^3*Y2+132*Y2^3*Y1+96*Y2^5*Y1
+1920*Y1^4*Y2^2+480*Y1^2*Y2^4+896*Y1^2*Y2^3+72*Y1^2*Y2+36*Y2^2*Y1+25*Y1*
Y2+512*Y1^6+640*Y1^5+48*Y1^3+22*Y1^2+200*Y1^4+7*Y2^2
+24*Y2^5+8*Y2^6+6*Y2^3+18*Y2^4-3*Y1-2*Y2;

temp2 := -3328*Y1^4*Y2-3456*Y1^3*Y2^2-3072*Y1^5*Y2-2560*Y1^3*Y2^3-
724*Y1^2*Y2^2-464*Y2^4*Y1-880*Y1^3*Y2-264*Y2^3*Y1-192*Y2^5*Y1
-3840*Y1^4*Y2^2-960*Y1^2*Y2^4-1792*Y1^2*Y2^3-120*Y1^2*Y2-60*Y2^2*Y1-
39*Y1*Y2-1024*Y1^6-1280*Y1^5-80*Y1^3-34*Y1^2-400*Y1^4-11*Y2^2
-48*Y2^5-16*Y2^6-10*Y2^3-36*Y2^4+6*Y1+4*Y2;

Y1 := temp1;
Y2 := temp2;
return(Y1, Y2);
end proc;
```

If one were to take values returned by this program and run them through the mappings provided by the function $\psi$ so that

$$\begin{bmatrix} V1 \\ V2 \end{bmatrix} = \psi\left(\begin{bmatrix} Y1 \\ Y2 \end{bmatrix}\right) = \begin{bmatrix} 5*Y1 + 3*Y2 + 8*Y1^2 + 8*Y1Y2 + 2*Y2^2 \\ -2*Y1 - Y2 \end{bmatrix}$$

the same output as the original function would be obtained.

**Some Properties of the Algorithm**

We construct procedures $KG$, $F_\varphi$, $E_\varphi$, $D_\varphi$ where:

I. *KG:* $\varphi \leftarrow$ Aut $\mathbb{Z}^n$ = the group of all polynomial mappings from $\mathbb{Z}^n$ to itself with an inverse $\psi$ of the same form. This plays the role of the security parameter in this scheme
II. $E_\varphi(P)$ is a new program with directed graph, state space, input, and output all isomorphic to those of the original program.
III. $E_\phi: \mathbb{Z}^k \rightarrow \mathbb{Z}^n$ is an asymmetric encryption function.
IV. $D_\varphi: \mathbb{Z}^n \rightarrow \mathbb{Z}^l$ is an asymmetric decryption function.
V. $D_\varphi(E_\varphi(P)(E_\varphi(u))) = P(u)$, i.e. the transformed program correctly processes encrypted data.

Some of the properties of the scheme are:

i. Let $\varphi(x) = (\varphi_1(x), \ldots, \varphi_n(x))$ be the component polynomials of the map $\varphi$, with $x = x_1, \ldots, x_n$. Assume that each polynomial $\varphi_1(x), \ldots, \varphi_n(x)$ has been written in standard (dense) form. Let $d(\varphi)$ denote the maximum degree of $\varphi_1(x), \ldots, \varphi_n(x)$. Let $|\varphi|$ denote the maximum absolute value of any coefficient of $\varphi_1(x),\ldots,\varphi_n(x)$. Let $m(\varphi)$ denote the maximum number of monomials of $\varphi_1(x),\ldots,\varphi_n(x)$. Then

$$|E\varphi(u)| \leq |\varphi|\, m(\varphi)\, (|RG|\vee|u|)^{d(\varphi)}$$

where $|RG|$ denotes the maximum output of the random number generator used in the encryption procedure. Consequently, an input $u$ to the program $P$, consisting of $m$ $b$-bit numbers, will be encrypted as $n$ $B$-bit numbers with $B = d(\varphi)b + \log(|\varphi|\, m(\varphi))$, assuming that the random number generator also produces $b$-bit numbers.

ii. If $b = 32$, i.e. the program $P$ uses ordinary 32-bit integer variables, and $\varphi$ is quadratic with each component consisting of at most 4 monomials, each of whose coefficients was bounded by $2^{32}$, then the transformed program would require the use of 98-bit integers.

iii. Using any measure of complexity, the complexity of transformations $E\varphi$, $D\varphi$ are bounded by the complexity of $\varphi$ and are independent of the complexity of $P$.

iv. The complexity of the transformed program is expressible in terms of the complexity of the original program, along with elementary measures of the security parameter $\varphi$.

v. Under simple and general conditions, the complexity of the transformed program will grow by approximately 2 or 3 orders of magnitude over the complexity of the original program, which is significantly less than other existing schemes.

vi. The encryption functions $E\varphi$, $D\varphi$ are a new type of multivariate encryption, although are basically just the methods of Moh, Patarin, Masumoto, Imai, et.al. (see [Mo],[P], [Mi]), in the setting of the integers rather than a finite field. To cryptoanalyze $E\varphi$ one must solve $n$ nonlinear equations in $k$ unknowns with integer values. There are several techniques for attacking this problem, the most common being Grobner basis methods. The complexity of these methods is exponential in $n$ (see [CLO],[LA], [LL],[Y]). Consequently the difficulty of cryptoanalyzing this encryption algorithm is exponential in $n$.

The basis of our method is the exploitation of nonlinear algebraic automorphisms of $\mathbb{Z}^n$. These objects can be thought of in several ways. The first is as a system of polynomial functions in several variables with integer coefficients. Inverting such an automorphism is equivalent to finding rational solutions to the system of equations with integral variables and coefficients. This is a hard thing to do, both in the general, theoretical sense and in any practical, computational sense, at least when dealing with more than a few variables. Another way to think of these automorphisms is as random curvilinear coordinate systems for vector spaces that keep integer lattices fixed. Using the first conceptualization of these automorphisms gives a new form of multivariate encryption. Using the second conceptualization yields a software obfuscation algorithm.

**Constructing automorphisms, the key generation algorithm**

To quantify matters, let $S$ be a polynomial mapping of $\mathbb{Z}^n$, $S(x) = (S_1(x),\ldots, S_n(x))$ with $x = x_1, \ldots x_n$. Assume that each polynomial $S_1(x),\ldots,S_n(x)$ has been written in standard, dense, form. Let $d(S)$ denote the maximum degree of $S_1(x),\ldots,S_n(x)$. Let $|S|$ denote the maximum absolute value of any coefficient of $S_1(x),…,S_n(x)$. Let $m(S)$ denote the maximum number of monomials of $S_1(x),…,S_n(x)$. Let $\bar{m}(S)$ denote the average number of monomials of $S_1(x),…,S_n(x)$. We shall display an algorithm which furnishes automorphism pairs that are constrained with respect to all these measures of complexity.

When constructing polynomial automorphisms of $\mathbb{Z}^n$, one thing that quickly becomes apparent is that the inverses are generically very complex (see [K] and [E]). We describe a way to generate nonlinear polynomial automorphisms of $\mathbb{Z}^n$ that allows for control on the complexity of the inverses. Not surprisingly, these automorphisms are *tame*, that is, they are constructed by composing a sequence of automorphisms which alternate between *affine* (linear with offset) and *triangular*. Triangular automorphisms are of the form

$$T(x_1,…, x_n) = (x_1 + f_1(x_2,…, x_n), …, x_i + f_i(x_{i+1},…,x_n),…, x_n + f_n)$$

where each of the functions $f_i$ is a polynomial in the indicated variables. Such a transformation is easily inverted with recursion. If $y = T(x)$, then one can solve for each of the variables $x_i$ by starting with the last and going in reverse:

$$x_n = y_n - f_n$$
$$x_{n-1} = y_{n-1} - f_{n-1}(x_n) = y_{n-1} - f_{n-1}(y_n - f_n)$$
$$x_{n-2} = y_{n-2} - f_{n-2}(y_{n-1} - f_{n-1}(y_n - f_n), y_n - f_n)$$
$$…$$

From this recursion we obtain $x_i = g_i(y_i,…,y_n)$ with

$$g_n(y) = y_n - f_n$$

and

$$g_i(y) = y_i - f_i(g_{i+1}(y),…,g_n(y)) \text{ for } i = n-1, n-2,…,1.$$

Then

$$T^{-1}(y) = (g_1(y),…, g_n(y)).$$

An important property of triangular automorphisms is that any measure of the inverse transformation typically grows very quickly with the number of variables $n$. For instance, the generic situation is that

$$\text{degree}(g_i) = \text{degree}(g_{i+1})^{\text{degree}(f_i)}$$

Consequently, if a generic triangular automorphism $T$ consisted of quadratic terms, the inverse would have terms of degree $2^{2^{n-1}}$. Even a few variables would result in inverses with enormous degree and complexity. If such transformations were used as factors of tame transformations and applied to program rewrites, things would quickly become untenable. To ameliorate this, suppose that the indices $\{1,...,n\}$ are subdivided into two disjoint sets $E_1$, $E_2$. Suppose that for $i \in E_2$ we have $f_i = 0$. In this case $g_i(y) = y_i$. Suppose also for $i \in E_1$ the polynomial $f_i$ is dependent only on variables whose indices are in $E_2$. Then, for those indices, $g_i(y) = y_i - f_i(y_{i+1},...,y_n)$. We call triangular transformations obtained in this way *segmented by the partition $E_1$, $E_2$*. It is clear that for such transformations

$$d(T^{-1}) = d(T), |T^{-1}| = |T|, m(T^{-1}) = m(T), \overline{m}(T^{-1}) = \overline{m}(T).$$

We will generate random sparse small segmented triangular nonlinear transformations as follows:

The input to the procedure are parameters $\beta, d, \mu, \bar{\mu}$.

I. Let $E_1 = \{1,..., \lceil n/2 \rceil\}$, $E_2 = \{\lceil n/2 \rceil + 1, ..., n\}$.
II. For $i \in E_2$ set $f_i = 0$.
III. For $i \in E_1$ randomly generate a number $\mu_i$ in the range $[1, \mu]$ with mean $E[\mu_i] = \bar{\mu}$. Generate $2\mu_i$ monomials $m_j$ of degree $\leq d$ in the variables indexed by $E_2$ with coefficients bounded by in absolute value. Set $f_i = \sum_j m_j$. Repeat this process until a nonlinear, nonhomogenous $f_i$ is obtained.
IV. Set $T(x) = (x_1 + f_1(x_2,..., x_n), ..., x_i + f_i(x_{i+1},...,x_n),..., x_n + f_n)$.

Then it is clear from the construction that:

i. $d(T^{-1})$, $d(T) \leq d$.
ii. $|T^{-1}|$, $|T| \leq \beta$.
iii. $m(T^{-1})$, $m(T) \leq \mu$.
iv. $E[\overline{m}(T^{-1})\}, E[\overline{m}(T)] \leq \bar{\mu}$.

Once we have constructed triangular transformations, we generate affine automorphisms of $\mathbb{Z}^n$, then compose these automorphisms to produce $\varphi$. These too can result in inverses with large coefficients. To prevent this, we create affine transformations as follows:

Generate small sparse random affine transformations by first generating small sparse unimodular matrices. We will do this by first generating a block diagonal matrix consisting of small unimodular matrices of sizes 1 by 1 and 2 by 2. We will then multiply the resulting matrix on left and right by permutation matrices.

I. To generate small random unimodular 2 by 2 matrices, with norm bounded by $\beta$ and determinant $\delta = \pm 1$. Proceed as follows:

1. Pick two random coprime numbers $x_{11}$, $x_{12}$ both with absolute value $\leq \beta$.

2. Solve the equation $x\,x_{11} - y\,x_{12} = \delta$. Since $x_{11}$ and $x_{12}$ are coprime, this equation is solvable. Find the solution with minimal absolute value. Then $|x|, |y| \leq \beta$.
3. Then $M = \begin{bmatrix} x_{11} & x_{12} \\ x & y \end{bmatrix}$ is unimodular with $|M|, |M^{-1}| \leq \beta$.

II. To generate a sparse, small $n$ by $n$ unimodular block diagonal matrix, start with inputs $\beta \geq 1$, $\alpha$ in the range $[1, n]$. Then:

1. Generate random numbers $\delta_i = \pm 1$, for $i = 1, ..., n - 2\alpha + 1$.
2. Use I. to generate $\alpha$ distinct 2 by 2 unimodular matrices $M_1, ..., M_\alpha$ with $|M_i|, |M_i^{-1}| \leq \beta$ and $\det(M_i) = \delta_i$.
3. Define $\Delta$ by using the matrices $M_1, ..., M_\alpha$ along with 1 by 1 matrices $[\delta_i]$ for $i = \alpha + 1, ... n - 2\alpha + 1$ on the diagonal of $\Delta$

The resulting matrix $\Delta$ has the properties:

i. Determinant $\delta(\Delta) = \prod \delta_i$,
ii. $|\Delta|, |\Delta^{-1}| \leq \beta$,
iii. $m(\Delta), m(\Delta^{-1}) \leq 2$,
iv. $\bar{m}(\Delta) = \bar{m}(\Delta^{-1}) = 1 + 2\alpha/n$.

III. To generate an arbitrary small, sparse, unimodular $n$ by $n$ matrices. Start with parameters $\beta \geq 1$, $\bar{\mu}$ in the range $(1,2)$.

1. Pick a random integer $\alpha$ between 1 and $n$ with expected value $n(\bar{\mu}-1)/2$.
2. Use II to generate a random unimodular block diagonal matrix $\Delta$.
3. Generate two random $n$ by $n$ permutation matrices $P_1, P_2$,
4. Let $M = P_1 \Delta P_2$

The resulting matrix $M$ has the properties:

i. $\delta(M) = \pm 1$
ii. $|M|, |M^{-1}| \leq \beta$,
iii. $m(M), m(M^{-1}) \leq 2$,
iv. $E[\bar{m}(M)] = E[\bar{m}(M^{-1})] = \bar{\mu}$.

IV. To generate a small sparse affine transformation $A$, start with parameters $\beta \geq 1$, $\bar{\mu}$, and $\mu$. Let $v = \lfloor \log \mu \rfloor$.
1. Use III to generate matrices $M_1, ..., M_v$ with parameters $(\beta/\mu)^{1/v}$ and $\bar{\mu}^{1/v}$.
2. Let $M = M_1 \times ... \times M_v$.

The resulting affine transformation has the properties:

i. $|A|, |A^{-1}| \leq \beta$,
ii. $E[\bar{m}(A)] = E[\bar{m}(A^{-1})] \leq \bar{\mu}$.
iii. $m(A), m(A^{-1}) \leq \mu$.

To generate tame automorphisms of $\mathbb{Z}^n$, start with parameters $b, d, m, \bar{m}, k$.

Choose $\delta(1),\ldots,\delta(k)$ so $\Pi = \Pi_{i \leq k}\, \delta(i) \leq d$. Let $\Delta(i) = \delta(1)\cdot\ldots\cdot\delta(i)$, $\Sigma = 1+\Delta(1)+\ldots+\Delta(k-1)$.

Choose $\mu_t, \mu_a$ so that $(\mu_a\mu_t)^{\Sigma}(\mu_a)^{\Delta(k)} \leq m$.

Choose $\beta_t, \beta_a$ so that $(\beta_a\mu_a\beta_t\mu_t)^{\Sigma}(\beta_a\mu_a)^{\Delta(k)} \leq b$.

Choose $\bar{\mu}_t, \bar{\mu}_a$ so that $(\bar{\mu}_a\bar{\mu}_t)^{\Sigma}(\bar{\mu}_a)^{\Delta(k)} \leq \bar{m}$.

I. Use the affine generation algorithm with parameters $\beta_a, \mu_a, \bar{\mu}_a$ to generate a sequence of sparse, small affine transformations $A_0,\ldots,A_k$ with the properties that:

   i. $|A_i|, |A_i^{-1}| \leq \beta_a$,
   ii. $E[\bar{m}(A_i)] = E[\bar{m}(A_i^{-1})] \leq \bar{\mu}_a$.
   iii. $m(A_i) \leq \mu_a$.

II. Use the triangular generation algorithm with parameters $\beta_t, \delta(i), \mu_t, \bar{\mu}_t$ sequence of small sparse triangular transformations $T_1,\ldots,T_k$ with the properties that:

   i. $|T_i|, |T_i^{-1}| \leq \beta_t$,
   ii. $E[\bar{m}(T_i)] = E[\bar{m}(T_i^{-1})] \leq \bar{\mu}_t$.
   iii. $m(T_i), m(T_i^{-1}) \leq \mu_t$.
   iv. $d(T_i), d(T_i^{-1}) \leq \delta(i)$.

Let $\varphi = A_0 \circ T_1 \ldots \circ T_k \circ A_k$.

It is easy to see that

$$|\varphi| \leq |A_0|m(A_0)|T_1|m(T_1)(|A_1|m(A_1))^{\Delta(1)}(|T_2|m(T_2))^{\Delta(1)}\ldots(|T_k|m(T_k))^{\Delta(k-1)}|A_k|^{\Delta(k)}$$
$$\leq (\beta_a\mu_a\beta_t\mu_t)(\beta_a\mu_a\beta_t\mu_t)^{\Delta(1)}(\beta_a\mu_a\beta_t\mu_t)^{\Delta(2)}\ldots (\beta_a\mu_a\beta_t\mu_t)^{\Delta(k-1)}(\beta_a\mu_a)^{\Delta(k)}$$
$$=(\beta_a\mu_a\beta_t\mu_t)^{\Sigma}(\beta_a\mu_a)^{\Delta(k)}$$

$$m(S) \leq m(A_0)m(T_1)m(A_1)^{\Delta(1)}m(T_2)^{\Delta(1)}m(A_2)^{\Delta(2)}\ldots m(T_k)^{\Delta(k-1)}m(A_k)^{\Delta(k)} \leq (\mu_a\mu_t)^{\Sigma}(\mu_a)^{\Sigma(k)}$$

Consequently $\varphi$ has the properties that:

   i. $d(\varphi) \leq \Pi$.
   ii. $|\varphi| \leq (\beta_a\mu_a\beta_t\mu_t)^{\Sigma}(\beta_a\mu_a)^{\Delta(k)}$

   iii.    $m(\varphi) \leq (\mu_a \mu_t)^{\Sigma} (\mu_a)^{\Delta(k)}$

   iv.    $E[\bar{m}(\varphi)] \leq (\bar{\mu}_a \bar{\mu}_t)^{\Sigma} (\bar{\mu}_a)^{\Delta(k)}$

If $k = 1$, the bounds above correspond to:

   i.    $d(S) \leq \delta(1)$.

   ii.    $|S|, |S^{-1}| \leq \beta_t \mu_t (\beta_a \mu_a)^{\delta(1)+1}$

   iii.    $m(S), m(S^{-1}) \leq \mu_t \mu_a^{\delta(1)+1}$

   iv.    $E[\bar{m}(S)], E[\bar{m}(S^{-1})] \leq \bar{\mu}_t \bar{\mu}_a^{\delta(1)+1}$

We obtain:

**Theorem:** *The algorithm, with inputs $n, d, b, m, \bar{m}$ described above yields an automorphisms $\varphi$ of $\mathbb{Z}^n$ such that:*

1. $d(\varphi), d(\varphi^{-1}) \leq d$.
2. $|\varphi|, |\varphi^{-1}| \leq b$.
3. $m(\varphi), m(\varphi^{-1}) \leq m$
4. $E[\bar{m}(\varphi)], E[\bar{m}(\varphi^{-1})] \leq \bar{m}$.

Examples:

If $n=2$, $b=3$, $d=2$, $m=5, \bar{m} = 5$ a typical automorphism pair is

$$\phi := \begin{bmatrix} -X_2 + X_1^2 + 2 X_1 X_2 + X_2^2 - 2 X_1 \\ -2 X_2 + X_1^2 + 2 X_1 X_2 + X_2^2 - 3 X_1 \end{bmatrix}$$

$$\psi := \begin{bmatrix} -2 Y_1 + Y_1^2 - 2 Y_1 Y_2 + Y_2^2 + Y_2 \\ 3 Y_1 - Y_1^2 + 2 Y_1 Y_2 - Y_2^2 - 2 Y_2 \end{bmatrix}$$

If $n = 2$, $b = 10^{12}$, $d=2$, $m=5, \bar{m} = 5$ a typical automorphism pair is

$$\phi := \begin{bmatrix} -2331187 X_1 + 2246855 X_2 + 14309229798 X_1^2 - 27583166484 X_1 X_2 + 13292662918 X_2^2 \\ -6593429 X_1 + 6354908 X_2 + 40471627563 X_1^2 - 78015075354 X_1 X_2 + 37596412283 X_2^2 \end{bmatrix}$$

$$\psi := \begin{bmatrix} -6354908 Y_1 + 2246855 Y_2 - 38201180309 Y_1^2 + 27012971828 Y_1 Y_2 - 4775380244 Y_2^2 \\ -6593429 Y_1 + 2331187 Y_2 - 39635004771 Y_1^2 + 28026863532 Y_1 Y_2 - 4954617036 Y_2^2 \end{bmatrix}$$

**Variations on the key generation system**

Our key generation procedure is easily generalized, for example one could require that for indices $i \in E_2$ the polynomial $f_i$ is a constant (rather than 0). Another generalization is that for indices $i \in E_1$ *all but one* of the monomials comprising the polynomial $f_i$ depends only on variables whose indices are in $E_2$. Still another generalization is to compose more than two affine transformations with one triangular transformation, for example one could use $\varphi = A_0 \circ T_1 \circ A_1 \circ T_2 \circ A_2$. Many variations are possible, allowing for the generation of tame automorphisms whose inverses are difficult to discover but feasible to use.

**Encryption and Decryption**

Having constructed an algebraic autorphism $\varphi$ of $\mathbb{Z}^n$, perform encryption as follows.:

**Encryption Algorithm (version 0)**

Encrypt vectors of integers of length $n$ by applying the nonlinear automorphism $\varphi$ to this vector to obtain the ciphertext, a vector of integers of length $n$ so $E_\varphi(u) = \varphi(u)$. Decryption uses the inverse transformation $\psi = \varphi^{-1}$ so $D_\varphi(u) = \psi(u)$.

This algorithm furnishes public key encryption. One can publish $\varphi$ and the random vector generator (the public keyed encryption algorithm) and keep $\psi$ private.

The strength of this encryption algorithm rests on the difficulty of solving systems of nonlinear algebraic equations. To break the encryption system, one must be able to solve, in $x$, the system of equations corresponding to the relation $\varphi(x) = y$, where $y$ is the ciphertext and $x$ is the plaintext. Here $x$ and $y$ can be considered as vectors of length $n$ with integer values. There are several methods of solving these systems, the most important relies on Grobner bases to perform elimination theory. In all cases, these algorithms are exponential in $n$ (see [LL]). An alternative attack is to start with knowledge of the generation algorithm and examine all possible component mappings. It is easy to bound the coefficients of these components using the size of the coefficients occurring in automorphism appearing in the encryption. However, is also easy to prove that the size of the search space grows exponentially with the size of the coefficients and super-exponentially with the number of variables.

There are several methods of adding random number generation to this encryption algorithm that are analogous to PKE algorithms employing finite fields rather than integers and are also compatible with the program modification procedure of our scheme. Two such are the following.

**Encryption Algorithm (version 1)**

Along with the automorphism $\varphi$ of $\mathbb{Z}^n$, start with some $m < n$, and a random number generator $G$. For $u \in \mathbb{Z}^m$, let $E_\varphi(u) = \varphi(u, g)$, where $g \leftarrow G^{(n-m)}$ denotes a random vector of length $n-m$ formed by using $G$ repeatedly. This method relies on a careful choice of $\varphi$ to insure that sufficient mixing of the input components occur, otherwise some components of $E_\varphi(u)$ may not exhibit any mixing with $g$. Decryption is $D_\varphi(u,g) = \psi(u,g)$, followed by truncation of the random component $g$.

**Encryption Algorithm (version 2)**

Along with the automorphism $\varphi$ of $\mathbb{Z}^n$, some $m < n$, and a random number generator $G$, start with $h: \mathbb{Z}^{n-m} \to \mathbb{Z}^m$, $H: \mathbb{Z}^{n-m} \to \mathbb{Z}^{n-m}$ polynomial mappings with $H$ an automorphism. Define $E_\varphi(u) = \varphi(u+h(g), H(g))$ where $g \leftarrow G^{(n-m)}$. Decryption is performed by first applying $\psi$ to obtain $(v_1, v_2)$ with $v_1 \in \mathbb{Z}^m$, $v_2 \in \mathbb{Z}^{n-m}$, followed by $(v_1, v_2) \to (v_1 - h(H^{-1}(v_2)), v_2)$, followed by truncation.

This version of encryption guarantees mixing of all the components of the plaintext message with random numbers.

**Encrypting the Program**

An automorphism of $\mathbb{Z}^n$ can be used to perform a type of encryption of the program. This is a new kind of mathematical algorithm that transforms any set of instructions of any straight line computer program.

We start with the description of a computer program $P$ as above and an automorphism $\varphi$ of $\mathbb{Z}^n$, with inverse $\psi$.

Perform a rewrite of expressions in the instruction set of $P$ as follows:

Define a new set of variables $y = (y_1,\ldots,y_n)$ by $y = \varphi(x)$, so $x = \psi(y)$. Now rewrite $P$ following the procedure:

  I. Replace the computation instruction $x \leftarrow f_\alpha(x)$ by the instruction $y \leftarrow F_\alpha(y)$, where $F_\alpha(y)$ is a polynomial obtained from $\psi(f_\alpha(\varphi(y)))$ by expanding and simplifying the algebraic expression. These instructions are equivalent assuming the relationship $y = \varphi(x)$, $x = \psi(y)$. However trying to deduce $f_\alpha(x)$ from $\psi(f_\alpha(\varphi(y)))$ without knowing $\varphi$ and $\psi$ is generally a difficult problem. Note: this may require the introduction of dummy variables, as in the example above.

  II. Replace the operations $x \leftarrow in(u)$, $v \leftarrow out(x)$ by the operations $y \leftarrow \varphi(in(u))$, $v \leftarrow out(\psi(y))$.

  III. Replace the variables $x_1,\ldots,x_n$ by the variables $y_1,\ldots, y_n$.

The resulting program is denoted $E_\varphi(P)$. By construction $P$ and $E_\varphi(P)$, followed by $D_\varphi(v)$ have the same input-output relationships.

**Performance metrics**

Let $|P|$ (respectively $m(P)$, $d(P)$) denote the maximum of $|f|$ (resp. $m(f)$, $d(f)$) for any instruction $f$ appearing in $P$. The complexity of the encrypted $E_\varphi(P)$ is bounded by the complexities of $P$, $\varphi$, $\psi$ by the following

**Theorem:**

$$|E_\varphi(P)| \leq |\varphi| m(\varphi) m(P)^{d(\varphi)} |P|^{d(\varphi)} |\psi|^{d(P)d(\varphi)}$$

$$d(E_\varphi(P)) \leq d(\varphi) d(P) d(\psi)$$

$$m(E_\varphi(P)) \leq m(\varphi) m(P)^{d(\varphi)} m(\psi)^{d(P)d(\varphi)}$$

Example: Under simple conditions a program $P$ would have $d(P) = 2$ and $m(P) = 2$. If $\varphi$ and $\psi = \varphi^{-1}$ are both quadratic with $m(\varphi) = m(\psi) = 2$, then $d(F_\varphi(P)) \leq 8$ and $m(F_\varphi(P)) \leq 128$. Therefore the complexity of the transformed program would only grow by approximately 2 or 3 orders of magnitude.

**Program modification**

There are several variants of the procedure $F_\varphi(P)$ depending on what randomization was used to obtain the probabilistic encryption algorithm $E_\varphi$.

**FHE Version 0**

In the case that the desired encryption algorithm is of the form $E_\varphi(u) = \varphi(u)$, i.e. deterministic encryption, proceed as follows:

Set $n = m$. Generate an automorphism $\varphi$ of $\mathbb{Z}^n$ and use it to Encrypt $P$. Remove the input operation $y \leftarrow \varphi(in(u))$ and replace with the new input operation $y \leftarrow w$. Remove the output operation $v \leftarrow out(\psi(y))$ and replace it with new output operation $z \leftarrow y$. The resulting program is $E_\varphi(P)$. The encryption process is the mapping $u \to \varphi(in(u))$. The decryption process is the mapping $z \to out(\psi(z))$.

**FHE Version 1**

In the case that the desired encryption algorithm is version I, proceed as follows: Initially the state variables of $P$ are $x' = (x_1, \ldots, x_m)$, the input variables are $u' = (u_1, \ldots, u_k)$. Let $x'' = (x_{m+1}, \ldots, x_n)$, $u'' = (u_{k+1}, \ldots, u_{k+n-m})$. Let $x = (x', x'') = (x_1, \ldots, x_n)$ be the new, expanded, set of state variables. Let $u = (u', u'') = (u_1, \ldots, u_{k+n-m})$ be the new, expanded set of input variables. Augment the original input operation $x' \leftarrow in(u')$ by $(x', x'') \leftarrow (in(u'), u'')$. Call this new program $P'$. Since $x''$ does not appear in any of the instructions used by $P$, $u''$ cannot effect the output of $P'$. Choose an automorphism $\varphi$ of $\mathbb{Z}^n$ and use it to encrypt $P'$ as above, obtaining $E_\varphi(P')$. The new state variables are $y = \varphi(x)$. Remove the input operation of $E_\varphi(P')$ and replace it by $y \leftarrow w$.

Remove the output operation $v \leftarrow out(\psi(y))$ of $E_\varphi(P')$ and replace it with new output operations $z \leftarrow y$. The resulting program is $F_\varphi(P)$. The encryption process is then $E_\varphi(u) = \varphi(in(u), g)$. The decryption process is the mapping $D_\varphi(z) = out(\psi(z)_t)$, here the subscript refers to truncation. If $w = E_\varphi(u)$ is input to $F_\varphi(P)$, the value of the random input $g$ will not affect the output of $D_\varphi$, although the processing of the state variables of $F_\varphi(P)$ will be affected.

## FHE Version 2

In the case that the desired encryption algorithm is version 2, proceed as follows: Initially suppose that the state variables of $P$ are $x' = (x_1,\ldots, x_m)$, the input variables are $u' = (u_1,\ldots, u_k)$. Let $x'' = (x_{m+1},\ldots, x_n)$, $u'' = (u_{k+1},\ldots, u_{k+n-m})$. Let $x = (x', x'') = (x_1,\ldots, x_n)$ be the new, expanded set of state variables. Let $u = (u', u'') = (u_1,\ldots, u_{k+n-m})$ be the new, expanded set of input variables. Replace the original input operation $x' \leftarrow in(u')$ by $(x', x'') \leftarrow (in(u')+h(u''),H(u''))$. Immediately after this instruction, insert the instruction $x \leftarrow (x'-h(H^{-1}(x'')),K(x))$, here $K$ is any randomly chosen polynomial. Call this new program $P'$. Since $x''$ does not appear in any of the instructions used by $P$, $u''$ cannot effect the output of $P'$, and so neither can the choice of $K$. Choose an automorphism $\varphi$ of $\mathbb{Z}^n$ and use it to encrypt $P'$ as above. The new state variables are $y = \varphi(x)$. Remove the operation of $E_\varphi(P')$ corresponding to the input operation $(x', x'') \leftarrow (in(u') + h(u''), H(u''))$ and replace it by $y \leftarrow w$. Remove the output operation $v \leftarrow out(\psi(y))$ and replace it with new output operation $z \leftarrow y$. The resulting program is $\varphi(P)$. The encryption process is then $E_\varphi(u) = \varphi(in(u) + h(g), H(g))$. The decryption process is the mapping $D_\varphi(z) = out(\psi(z))$. If $w = E_\varphi(u)$ is input to $F_\varphi(P)$, the value of the random input $g$ will not affect the output of $D_\varphi$, although the processing of the state variables of $F_\varphi(P)$ will be affected.

One serendipitous feature of this program modification is that the calculations and decisions occurring during program execution are randomized and homogenized since every state variable is present and modified in every instruction. This acts as a defense against reverse engineering of the program by observing the program flow.

## Software Design Implications

The FHE scheme described in this paper is conceptually easy to implement: One starts with a program, identifies the variables in the program, generates an automorphism pair, and then rewrites the source code of the program. In practice, things are more complicated. Modern source code is modular and dependent on hidden libraries, definitions, etc. Realizing a piece of source code with the model assumed in this paper would require extensive analysis. Consequently, one would perform this encryption process in an incremental and adaptive fashion. In addition, performing encryption efficiently requires knowledge of the trade-offs between complexity and security that the process imposes. Judgment is required as to what trade-offs are acceptable, which frequently necessitates domain expertise. Finally, the size of integer variables and calculations in a modified program grows significantly compared to the original program, which requires additional measures to accommodate.

**Encrypting More General Programs**

One might consider applying this method to programs other than straight line programs. The following example illustrates the with the problem with this.

Start with the following program to compute 10!

```
fun5 := proc ( )
 local X1, X2, X3, temp1, temp2, temp3;
 X1 := 1;
 X2 := 1;
 temp1 := X1;
 temp2 := X2;
 while  X2 < 11 do
 temp1 := temp1·temp2;
 temp2 := X2 + 1;
 X1 := temp1;
 X2 := temp2;
 end do;
 return(X1);
 end proc;
```

Any attempt to encrypt this program using affine transformations will necessitate the rewrite of the term X2. This will reveal part of the function $\psi$, resulting in a compromise of the encryption.

This problem could be ameliorated by modifications of the program, for example by replacing the conditional statement by a more complex, but equivalent statement.

**Other Issues**

There are several variations on this encryption which should be explored. One is to finite fields and to employ this key generation algorithm or others that come from other forms of multivariate encryption. Another is to attempt to employ this method of encryption in the context of floating-point numbers. Still another is to examine the use of this in a rational number field setting.

**Conclusions**

This paper has described a method of performing Fully Homomorphic Encryption which results in modified programs that are fast enough to be employed practically. The scheme uses transformations of the program that are derived from the theory of affine automorphisms, a branch of algebraic geometry.

**Appendix: A demonstration of the algorithm with randomized state variables.**

A demonstration of the algorithm follows. Start with a simple program P which adds and multiplies two numbers: The transforming procedure will follow version 2 of the algorithm.

The initial program is:

```c
int main() //program P accepts two integers and returns their sum and product
{
        long x1=0,x2=0;
        scanf("%ld %ld",&x1,&x2);
        printf("%ld %ld\n",x1+x2,x1*x2);
        return(0);
}
```

Next, modify the program by including the procedure $(x', x'') \leftarrow (in(u')+h(u''), H(u''))$ to adjust the input into the program as described previously. The resulting program follows:

```c
int main()      //P', the first modification of the program
```

```c
{
    long x1,x2,x3,x4,u1,u2,g1,g2;
    //input
    scanf("%ld %ld",&u1,&u2);
    g1=rand();                          //
    g2=rand();                          // x <- (u+h(g),H(g))
    x1=u1+g1;                           // h(g) = (g1,g1*g2)
    x2=u2+g1*g2;                        // H(g) = (g1+g2,g2)
    x3=g1+g2;                           //
    x4=g2;                              //
    // end of input
    //
    // so  g2=x4
    // g1=x3-x4
    // u1=x1-g1=x1-(x3-x4) = x1-x3+x4
    // u2=x2-g1*g2=x2-(x3-x4)*x4=x2-x3*x4+x4*x4
    //
    x1=x1-x3+x4;                        //
    x2=x2-x3*x4+x4*x4;                  // x <- (x'-h(H^(-1)(x'')),K(x))
    x3=x1+2*x2+3*x3+4*x4;               // x'=(x1,x2)   x''=(x3,x4)
    x4=x1-6*x3;                         //K(x) = (x1+2*x2+3*x3+4*x4,x1-6*x3)
    printf("%ld %ld\n",x1+x2,x1*x2);
    return(0);
}
```

Introduce an automorphism $\varphi$ and an inverse $\psi$ of the state vectors by

$$\varphi = \begin{bmatrix} -4 - X2 - 2X2X4 - 2X3X4 \\ X4 - X1 - X3 + 1 \\ 4 + X2 + 2X2X4 + 2X3x_4 - X4 \\ 1 + X4 - X1 - 2X2 - 2X3 \end{bmatrix}$$

$$\psi = \begin{bmatrix} 1 - 2Y_2 + Y_4 - Y_1 - Y_3 \\ -Y_1 - 4 - 2Y_1Y_4 + 2Y_1Y_2 - 2Y_3Y_4 + 2Y_2Y_3 \\ Y_1 + 4 - Y_4 + Y_2 + 2Y_1Y_4 - 2Y_1Y_2 + 2Y_3Y_4 - 2Y_2Y_3 \\ -Y_1 - Y_3 \end{bmatrix}$$

Using this automorphism pair, encrypt the program P'. The resulting program follows:

```c
int main()    //E(P'), the encrypted version of P'
{
```

```
  long Y1 = 0, Y2 = 0, Y3 = 0, Y4 = 0, u1, u2, g1, g2;
  int Y1new, Y2new, Y3new, Y4new;

  scanf("%ld %ld",&u1,&u2);
  g1=rand();    g2=rand();
  Y1new = Y1;
  Y2new = -Y1-Y3-u1-g1+1+Y4-Y2;
  Y3new = Y3;
  Y4new = 1-Y1-Y3-u1-g1+2*Y4-2*Y2;
  Y1 = Y1new; Y2 = Y2new; Y3 = Y3new; Y4 = Y4new;
Y1new = -4-u2-g1*g2+2*u2*Y1+2*u2*Y3+2*g1*g2*Y1+2*g1*g2*Y3+2*Power(Y1,
2)+2*Y1*Y3+8*Y1+8*Y3-2*Y1*Y4-2*Y3*Y4+2*Y1*Y2+2*Y2*Y3+4*Power(Y1,
2)*Y4+8*Y1*Y4*Y3-4*Power(Y1, 2)*Y2-8*Y1*Y2*Y3+4*Power(Y3, 2)*Y4-
4*Y2*Power(Y3, 2);
  Y2new = -Y1+Y2-u2-g1*g2-4-2*Y1*Y4+2*Y1*Y2-2*Y3*Y4+2*Y2*Y3;
  Y3new = 4+u2+g1*g2-2*u2*Y1-2*u2*Y3-2*g1*g2*Y1-2*g1*g2*Y3-2*Power(Y1, 2)-
2*Y1*Y3-7*Y1-7*Y3+2*Y1*Y4+2*Y3*Y4-2*Y1*Y2-2*Y2*Y3-4*Power(Y1, 2)*Y4-
8*Y1*Y4*Y3+4*Power(Y1, 2)*Y2+8*Y1*Y2*Y3-4*Power(Y3,
2)*Y4+4*Y2*Power(Y3, 2);
  Y4new = -2*Y1+Y4-2*u2-2*g1*g2-8-4*Y1*Y4+4*Y1*Y2-4*Y3*Y4+4*Y2*Y3;
  Y1 = Y1new; Y2 = Y2new; Y3 = Y3new; Y4 = Y4new;
Y1new = -7*Y1+2*Y1*Y4-2*Y1*Y2+2*Y3*Y4-2*Y2*Y3-2*Power(Y1, 2)-2*Y1*Y3-
8*Y3-4*Power(Y1, 2)*Y4-8*Y1*Y4*Y3+4*Power(Y1, 2)*Y2+8*Y1*Y2*Y3-
4*Power(Y3, 2)*Y4+4*Y2*Power(Y3, 2)+2*g1*Y1+2*g1*Y3+2*g2*Y1+2*g2*Y3;
  Y2new = Y1+2*Y2-Y4+4+2*Y1*Y4-2*Y1*Y2+2*Y3*Y4-2*Y2*Y3-g1-g2;
  Y3new = -2*Y1*Y4+2*Y1*Y2-2*Y3*Y4+2*Y2*Y3+2*Power(Y1,
2)+2*Y1*Y3+8*Y1+9*Y3+4*Power(Y1, 2)*Y4+8*Y1*Y4*Y3-4*Power(Y1, 2)*Y2-
8*Y1*Y2*Y3+4*Power(Y3, 2)*Y4-4*Y2*Power(Y3, 2)-2*g1*Y1-2*g1*Y3-2*g2*Y1-
2*g2*Y3;
  Y4new = 2*Y1+2*Y2-Y4+8+4*Y1*Y4-4*Y1*Y2+4*Y3*Y4-4*Y2*Y3-2*g1-2*g2;
  Y1 = Y1new; Y2 = Y2new; Y3 = Y3new; Y4 = Y4new;
Y1new = Y1+2*Y1*Y4-2*Y1*Y2+2*Y3*Y4-2*Y2*Y3+2*g2*Y4-2*g2*Y2;
  Y2new = g2+Y2+Y1+Y3;
  Y3new = -Y1-2*Y1*Y4+2*Y1*Y2-2*Y3*Y4+2*Y2*Y3-g2-2*g2*Y4+2*g2*Y2;
  Y4new = g2+Y4+Y1+Y3;
  Y1 = Y1new; Y2 = Y2new; Y3 = Y3new; Y4 = Y4new;      Y1new = Y1;
  Y2new = 2*Y1+Y3+4+2*Y2-Y4+2*Y1*Y4-2*Y1*Y2+2*Y3*Y4-2*Y2*Y3;
  Y3new = Y3;
  Y4new = 4+2*Y1+Y3+Y2+2*Y1*Y4-2*Y1*Y2+2*Y3*Y4-2*Y2*Y3;
  Y1 = Y1new; Y2 = Y2new; Y3 = Y3new; Y4 = Y4new;
      Y1new = -4*Y3-12*Power(Y1, 2)*Y2*Y3-12*Y1*Y2*Power(Y3,
2)+12*Power(Y1, 2)*Y4*Y3+12*Y1*Y4*Power(Y3, 2)-8*Y1*Y4*Y3+8*Y1*Y2*Y3-
4*Power(Y1, 3)*Y2+10*Power(Y1, 2)*Y3+8*Y1*Power(Y3, 2)-4*Y2*Power(Y3,
3)+4*Power(Y1, 3)*Y4+4*Power(Y3, 3)*Y4-Y1*Y2+4*Power(Y1, 2)*Y2-
Y2*Y3+13*Y1*Y3+4*Y2*Power(Y3, 2)+Y1*Y4+Y3*Y4-4*Power(Y1, 2)*Y4-
```

```
      4*Power(Y3, 2)*Y4-3*Y1+4*Power(Y1, 3)+7*Power(Y3, 2)+2*Power(Y3,
    3)+6*Power(Y1, 2);
      Y2new = -4*Y1-4*Y3+Y2-2*Power(Y1, 2)*Y4+2*Power(Y1, 2)*Y2-2*Power(Y3,
    2)*Y4+2*Y2*Power(Y3, 2)-Power(Y3, 2)-4*Y1*Y4*Y3+4*Y1*Y2*Y3-3*Y1*Y3-
    2*Power(Y1, 2)-Y1*Y2+Y3*Y4-Y2*Y3+Y1*Y4;
      Y3new = 5*Y3+12*Power(Y1, 2)*Y2*Y3+12*Y1*Y2*Power(Y3, 2)-12*Power(Y1,
    2)*Y4*Y3-12*Y1*Y4*Power(Y3, 2)+8*Y1*Y4*Y3-8*Y1*Y2*Y3+4*Power(Y1, 3)*Y2-
    10*Power(Y1, 2)*Y3-8*Y1*Power(Y3, 2)+4*Y2*Power(Y3, 3)-4*Power(Y1, 3)*Y4-
    4*Power(Y3, 3)*Y4+Y1*Y2-4*Power(Y1, 2)*Y2+Y2*Y3-13*Y1*Y3-4*Y2*Power(Y3,
    2)-Y1*Y4-Y3*Y4+4*Power(Y1, 2)*Y4+4*Power(Y3, 2)*Y4+4*Y1-4*Power(Y1, 3)-
    7*Power(Y3, 2)-2*Power(Y3, 3)-6*Power(Y1, 2);
      Y4new = -8*Y1-8*Y3+Y4-4*Power(Y1, 2)*Y4+4*Power(Y1, 2)*Y2-4*Power(Y3,
    2)*Y4+4*Y2*Power(Y3, 2)-2*Power(Y3, 2)-8*Y1*Y4*Y3+8*Y1*Y2*Y3-6*Y1*Y3-
    4*Power(Y1, 2)-2*Y1*Y2+2*Y3*Y4-2*Y2*Y3+2*Y1*Y4;
     Y1 = Y1new; Y2 = Y2new; Y3 = Y3new; Y4 = Y4new;
            Y1new = 2*Y3+3*Y1-10*Power(Y3, 2)-20*Y1*Y3-10*Power(Y1, 2)-2*Y3*Y4-
    2*Y1*Y4;
      Y2new = 5*Y1+5*Y3+Y2+Y4-1;
      Y3new = -Y3-2*Y1+10*Power(Y3, 2)+20*Y1*Y3+10*Power(Y1,
    2)+2*Y3*Y4+2*Y1*Y4;
      Y4new = 10*Y1+10*Y3+3*Y4-2;
     Y1 = Y1new; Y2 = Y2new; Y3 = Y3new; Y4 = Y4new;      Y1new = -
    46*Y4+46*Y2+Y1+16*Power(Y2, 2)-30*Y2*Y4+14*Power(Y4,
    2)+48*Y1*Y4*Y2+12*Y1*Y2-24*Y1*Power(Y4, 2)-24*Y1*Power(Y2, 2)-
    24*Y3*Power(Y4, 2)-24*Power(Y2, 2)*Y3+48*Y3*Y4*Y2-12*Y1*Y4;
     Y2new = -23-7*Y2+7*Y4-6*Y1-12*Y1*Y4+12*Y1*Y2-12*Y3*Y4+12*Y2*Y3;
     Y3new = 23+39*Y4-38*Y2+Y3+6*Y1-16*Power(Y2, 2)+30*Y2*Y4-14*Power(Y4, 2)-
    48*Y1*Y4*Y2-24*Y1*Y2+24*Y1*Power(Y4, 2)+24*Y1*Power(Y2,
    2)+24*Y3*Power(Y4, 2)+24*Power(Y2, 2)*Y3+12*Y3*Y4-12*Y2*Y3-
    48*Y3*Y4*Y2+24*Y1*Y4;
     Y4new = -23-8*Y2+8*Y4-6*Y1-12*Y1*Y4+12*Y1*Y2-12*Y3*Y4+12*Y2*Y3;
     Y1 = Y1new; Y2 = Y2new; Y3 = Y3new; Y4 = Y4new;
            printf("%ld %ld\n",-3-2*Y2+Y4-2*Y1-Y3-2*Y1*Y4+2*Y1*Y2-
    2*Y3*Y4+2*Y2*Y3,-4+8*Y2+4*Y3-
    4*Y4+6*Y1*Y4*Y2+6*Y3*Y4*Y2+4*Y1*Y4*Y3-4*Y1*Y2*Y3+4*Y1*Y2-
    2*Y1^2*Y2+2*Y2*Y3+Y1*Y3-2*Y2*Y3^2-3*Y1*Y4-
    2*Y3*Y4+2*Y1^2*Y4+2*Y3^2*Y4+3*Y1+Y1^2-4*Y1*Y2^2-4*Y2^2*Y3-2*Y1*Y4^2-
    2*Y3*Y4^2);
      return(0);
    }
```

Remove the original input and output statements, which translate from the state variables of the program to the input and output variables and replace them with input and output statements written in the new state variables. This modification results in *F*(P).

int main()      //  F(P), another encypted version of P

```c
{
    long Y1, Y2, Y3, Y4, Y1new, Y2new, Y3new, Y4new;

    scanf("%ld %ld %ld %ld",&Y1,&Y2,Y3,Y4);
    Y1new = Y1;
    Y2new = 2*Y1+Y3+4+2*Y2-Y4+2*Y1*Y4-2*Y1*Y2+2*Y3*Y4-2*Y2*Y3;
    Y3new = Y3;
    Y4new = 4+2*Y1+Y3+Y2+2*Y1*Y4-2*Y1*Y2+2*Y3*Y4-2*Y2*Y3;
    Y1 = Y1new; Y2 = Y2new; Y3 = Y3new; Y4 = Y4new;
Y1new = -4*Y3-12*Power(Y1, 2)*Y2*Y3-12*Y1*Y2*Power(Y3, 2)+12*Power(Y1, 2)*Y4*Y3+12*Y1*Y4*Power(Y3, 2)-8*Y1*Y4*Y3+8*Y1*Y2*Y3-4*Power(Y1, 3)*Y2+10*Power(Y1, 2)*Y3+8*Y1*Power(Y3, 2)-4*Y2*Power(Y3, 3)+4*Power(Y1, 3)*Y4+4*Power(Y3, 3)*Y4-Y1*Y2+4*Power(Y1, 2)*Y2-Y2*Y3+13*Y1*Y3+4*Y2*Power(Y3, 2)+Y1*Y4+Y3*Y4-4*Power(Y1, 2)*Y4-4*Power(Y3, 2)*Y4-3*Y1+4*Power(Y1, 3)+7*Power(Y3, 2)+2*Power(Y3, 3)+6*Power(Y1, 2);
    Y2new = -4*Y1-4*Y3+Y2-2*Power(Y1, 2)*Y4+2*Power(Y1, 2)*Y2-2*Power(Y3, 2)*Y4+2*Y2*Power(Y3, 2)-Power(Y3, 2)-4*Y1*Y4*Y3+4*Y1*Y2*Y3-3*Y1*Y3-2*Power(Y1, 2)-Y1*Y2+Y3*Y4-Y2*Y3+Y1*Y4;
    Y3new = 5*Y3+12*Power(Y1, 2)*Y2*Y3+12*Y1*Y2*Power(Y3, 2)-12*Power(Y1, 2)*Y4*Y3-12*Y1*Y4*Power(Y3, 2)+8*Y1*Y4*Y3-8*Y1*Y2*Y3+4*Power(Y1, 3)*Y2-10*Power(Y1, 2)*Y3-8*Y1*Power(Y3, 2)+4*Y2*Power(Y3, 3)-4*Power(Y1, 3)*Y4-4*Power(Y3, 3)*Y4+Y1*Y2-4*Power(Y1, 2)*Y2+Y2*Y3-13*Y1*Y3-4*Y2*Power(Y3, 2)-Y1*Y4-Y3*Y4+4*Power(Y1, 2)*Y4+4*Power(Y3, 2)*Y4+4*Y1-4*Power(Y1, 3)-7*Power(Y3, 2)-2*Power(Y3, 3)-6*Power(Y1, 2);
    Y4new = -8*Y1-8*Y3+Y4-4*Power(Y1, 2)*Y4+4*Power(Y1, 2)*Y2-4*Power(Y3, 2)*Y4+4*Y2*Power(Y3, 2)-2*Power(Y3, 2)-8*Y1*Y4*Y3+8*Y1*Y2*Y3-6*Y1*Y3-4*Power(Y1, 2)-2*Y1*Y2+2*Y3*Y4-2*Y2*Y3+2*Y1*Y4;
    Y1 = Y1new; Y2 = Y2new; Y3 = Y3new; Y4 = Y4new;
    Y1new = 2*Y3+3*Y1-10*Power(Y3, 2)-20*Y1*Y3-10*Power(Y1, 2)-2*Y3*Y4-2*Y1*Y4;
    Y2new = 5*Y1+5*Y3+Y2+Y4-1;
    Y3new = -Y3-2*Y1+10*Power(Y3, 2)+20*Y1*Y3+10*Power(Y1, 2)+2*Y3*Y4+2*Y1*Y4;
    Y4new = 10*Y1+10*Y3+3*Y4-2;
    Y1 = Y1new; Y2 = Y2new; Y3 = Y3new; Y4 = Y4new;    Y1new = -46*Y4+46*Y2+Y1+16*Power(Y2, 2)-30*Y2*Y4+14*Power(Y4, 2)+48*Y1*Y4*Y2+12*Y1*Y2-24*Y1*Power(Y4, 2)-24*Y1*Power(Y2, 2)-24*Y3*Power(Y4, 2)-24*Power(Y2, 2)*Y3+48*Y3*Y4*Y2-12*Y1*Y4;
    Y2new = -23-7*Y2+7*Y4-6*Y1-12*Y1*Y4+12*Y1*Y2-12*Y3*Y4+12*Y2*Y3;
    Y3new = 23+39*Y4-38*Y2+Y3+6*Y1-16*Power(Y2, 2)+30*Y2*Y4-14*Power(Y4, 2)-48*Y1*Y4*Y2-24*Y1*Y2+24*Y1*Power(Y4, 2)+24*Y1*Power(Y2, 2)+24*Y3*Power(Y4, 2)+24*Power(Y2, 2)*Y3+12*Y3*Y4-12*Y2*Y3-48*Y3*Y4*Y2+24*Y1*Y4;
    Y4new = -23-8*Y2+8*Y4-6*Y1-12*Y1*Y4+12*Y1*Y2-12*Y3*Y4+12*Y2*Y3;
    Y1 = Y1new; Y2 = Y2new; Y3 = Y3new; Y4 = Y4new;
    printf("%ld %ld %ld %ld \n",Y1,Y2,Y3,Y4);
    return(0);
```

}

The encryption for input into this program is provided by the translation from the original input variables into the new state variables.

```c
int main()       //E, the encryption algorithm for F(P)
{
        long Y1, Y2, Y3, Y4, u1, u2, g1, g2;
        int Y1new, Y2new, Y3new, Y4new;

        scanf("%ld %ld",&u1,&u2);
        g1=rand();     g2=rand();
        Y1new = Y1;
        Y2new = -Y1-Y3-u1-g1+1+Y4-Y2;
        Y3new = Y3;
        Y4new = 1-Y1-Y3-u1-g1+2*Y4-2*Y2;
        Y1 = Y1new; Y2 = Y2new; Y3 = Y3new; Y4 = Y4new;
        Y1new = -4-u2-g1*g2+2*u2*Y1+2*u2*Y3+2*g1*g2*Y1+2*g1*g2*Y3+2*Power(Y1, 2)+2*Y1*Y3+8*Y1+8*Y3-2*Y1*Y4-2*Y3*Y4+2*Y1*Y2+2*Y2*Y3+4*Power(Y1, 2)*Y4+8*Y1*Y4*Y3-4*Power(Y1, 2)*Y2-8*Y1*Y2*Y3+4*Power(Y3, 2)*Y4-4*Y2*Power(Y3, 2);
        Y2new = -Y1+Y2-u2-g1*g2-4-2*Y1*Y4+2*Y1*Y2-2*Y3*Y4+2*Y2*Y3;
        Y3new = 4+u2+g1*g2-2*u2*Y1-2*u2*Y3-2*g1*g2*Y1-2*g1*g2*Y3-2*Power(Y1, 2)-2*Y1*Y3-7*Y1-7*Y3+2*Y1*Y4+2*Y3*Y4-2*Y1*Y2-2*Y2*Y3-4*Power(Y1, 2)*Y4-8*Y1*Y4*Y3+4*Power(Y1, 2)*Y2+8*Y1*Y2*Y3-4*Power(Y3, 2)*Y4+4*Y2*Power(Y3, 2);
        Y4new = -2*Y1+Y4-2*u2-2*g1*g2-8-4*Y1*Y4+4*Y1*Y2-4*Y3*Y4+4*Y2*Y3;
        Y1 = Y1new; Y2 = Y2new; Y3 = Y3new; Y4 = Y4new;      Y1new = -7*Y1+2*Y1*Y4-2*Y1*Y2+2*Y3*Y4-2*Y2*Y3-2*Power(Y1, 2)-2*Y1*Y3-8*Y3-4*Power(Y1, 2)*Y4-8*Y1*Y4*Y3+4*Power(Y1, 2)*Y2+8*Y1*Y2*Y3-4*Power(Y3, 2)*Y4+4*Y2*Power(Y3, 2)+2*g1*Y1+2*g1*Y3+2*g2*Y1+2*g2*Y3;
        Y2new = Y1+2*Y2-Y4+4+2*Y1*Y4-2*Y1*Y2+2*Y3*Y4-2*Y2*Y3-g1-g2;
        Y3new = -2*Y1*Y4+2*Y1*Y2-2*Y3*Y4+2*Y2*Y3+2*Power(Y1, 2)+2*Y1*Y3+8*Y1+9*Y3+4*Power(Y1, 2)*Y4+8*Y1*Y4*Y3-4*Power(Y1, 2)*Y2-8*Y1*Y2*Y3+4*Power(Y3, 2)*Y4-4*Y2*Power(Y3, 2)-2*g1*Y1-2*g1*Y3-2*g2*Y1-2*g2*Y3;
        Y4new = 2*Y1+2*Y2-Y4+8+4*Y1*Y4-4*Y1*Y2+4*Y3*Y4-4*Y2*Y3-2*g1-2*g2;
        Y1 = Y1new; Y2 = Y2new; Y3 = Y3new; Y4 = Y4new;
        Y1new = Y1+2*Y1*Y4-2*Y1*Y2+2*Y3*Y4-2*Y2*Y3+2*g2*Y4-2*g2*Y2;
        Y2new = g2+Y2+Y1+Y3;
        Y3new = -Y1-2*Y1*Y4+2*Y1*Y2-2*Y3*Y4+2*Y2*Y3-g2-2*g2*Y4+2*g2*Y2;
        Y4new = g2+Y4+Y1+Y3;
        Y1 = Y1new; Y2 = Y2new; Y3 = Y3new; Y4 = Y4new;
            printf("%ld %ld %ld %ld\n",Y1,Y2,Y3,Y4);
        return(0);
}
```

Cryptoanalysis of this encryption function requires inverting a system of four equations in four unknowns, doable, but nontrivial. This problem would be more difficult if more random numbers were used. The difficulty increases exponentially with the number of random numbers introduced.

The new decryption algorithm is

```c
int main()
{
        long Y1, Y2, Y3, Y4;

        scanf("%ld %ld %ld %ld",&Y1,&Y2,&Y3,&Y4);

printf("%ld %ld\n",-3-2*Y2+Y4-2*Y1-Y3-2*Y1*Y4+2*Y1*Y2-2*Y3*Y4+2*Y2*Y3,-4+8*Y2+4*Y3-4*Y4+6*Y1*Y4*Y2+6*Y3*Y4*Y2+4*Y1*Y4*Y3-4*Y1*Y2*Y3+4*Y1*Y2-2*Y1^2*Y2+2*Y2*Y3+Y1*Y3-2*Y2*Y3^2-3*Y1*Y4-2*Y3*Y4+2*Y1^2*Y4+2*Y3^2*Y4+3*Y1+Y1^2-4*Y1*Y2^2-4*Y2^2*Y3-2*Y1*Y4^2-2*Y3*Y4^2);
        return(0);
}
```

The resulting program is clearly more complex than the original, but by a manageable amount.